\documentclass{article}
\usepackage{spconf,amsmath,graphicx}
\usepackage{comment}
\usepackage{amssymb}
\usepackage{booktabs}
\usepackage{multirow}
\usepackage[pagebackref,breaklinks,colorlinks]{hyperref}
\usepackage{enumitem}
\setlist{nosep, leftmargin=14pt}

\usepackage{mwe} % to get dummy images

\title{TPOT: Topology Preserving Optimal Transport in Retinal Fundus Image Enhancement}
\name{Xuanzhao Dong$^{1}$ \quad Wenhui Zhu$^{1}$ \quad Xin Li$^{1}$ \quad Guoxin Sun$^{1}$ \quad Yi Su$^{2}$ \quad Oana M. Dumitrascu$^{3}$ \quad Yalin Wang$^{1}$ }
\address{$^{1}$  School of Computing and Augmented Intelligence, Arizona State University, AZ, USA \\
$^{2}$ Banner Alzheimer’s Institute, AZ, USA \quad $^{3}$ Department of Neurology, Mayo Clinic, AZ USA\\ 
}

\begin{document}
%\ninept
%
\maketitle
\begin{abstract}
Retinal fundus photography enhancement is important for diagnosing and monitoring retinal diseases. However, early approaches to retinal image enhancement, such as those based on Generative Adversarial Networks (GANs), often struggle to preserve the complex topological information of blood vessels, resulting in spurious or missing vessel structures. The persistence diagram, which captures topological features based on the persistence of topological structures under different filtrations, provides a promising way to represent the structure information. In this work, we propose a topology-preserving training paradigm that regularizes blood vessel structures by minimizing the differences of persistence diagrams. We call the resulting framework Topology Preserving Optimal Transport (TPOT). Experimental results on a large-scale dataset demonstrate the superiority of the proposed method compared to several state-of-the-art supervised and unsupervised techniques, both in terms of image quality and performance in the downstream blood vessel segmentation task. The code is available at \url{https://github.com/Retinal-Research/TPOT}.
\end{abstract}

\begin{keywords}
Retinal Fundus Photography, Image Denoising, GANs, Topology Preserving
\end{keywords}
\section{Introduction}
\label{sec:intro}
% background of CFP + related work - GAN based methods - OT-based algorithms. highlight the drawbacks of previous methods and its drawbacks in preserving vessel structure
% The application of topology loss in retinal images (mostly in vessel segmentation)
%contribution
\begin{figure*}[!t]
  \centering
  \includegraphics[width=0.9\textwidth]{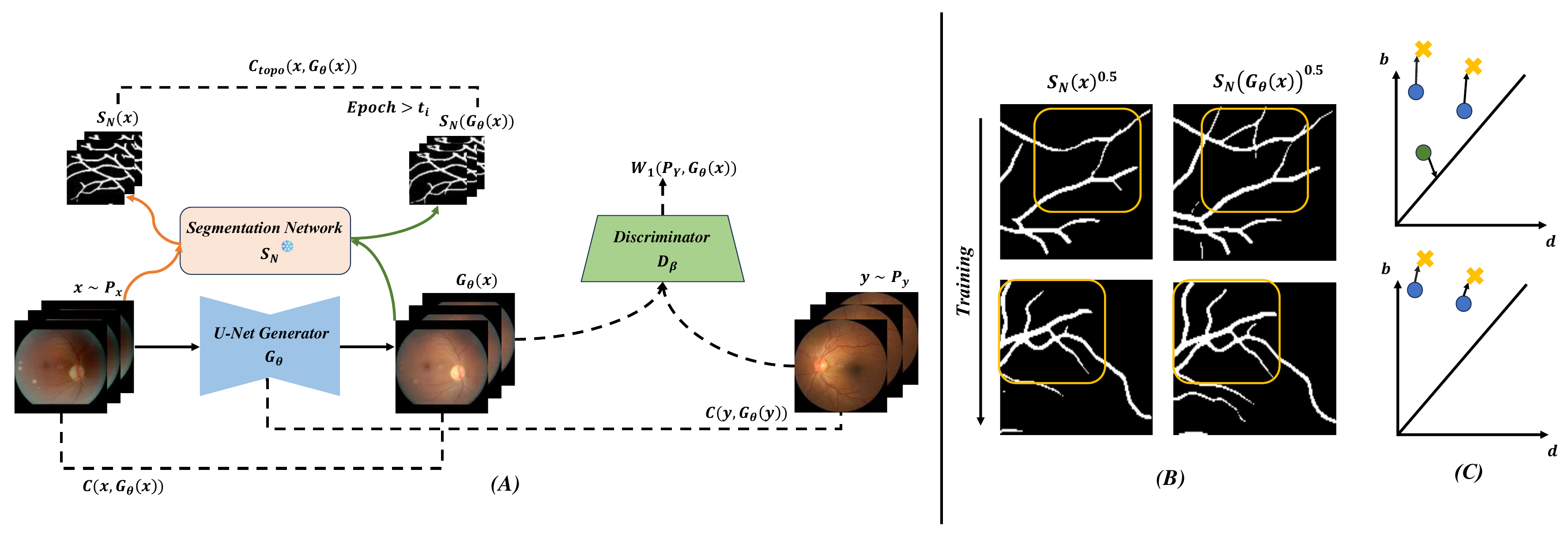}  % Replace with your figure file and remove the example
  \vspace{-0.4cm}
  \caption{\textbf{(A)} illustrates the TPOT framework. The structures of $G_\theta$ and $D_\beta$ follows the design outlined in~\cite{zhu2023otre,zhu2023optimal}, and $S_N$ follows the design presented in~\cite{zhou2021study}. \textbf{(B)} represents the changes in segmentation masks of the images during training. The orange box highlights regions with complex topological structures. \textbf{(C)} provides an example of how our topology-preserving regularization operates. The yellow crosses and blue dots represent corresponding persistent features in $S_N(\mathbf{x})$ and $S_N(G_\theta(\mathbf{x}))$, respectively. The green dots indicate persistent noise features that must be removed during training. The model penalizes the differences between corresponding points to encourage topological consistency. Further details are discussed in Sec.~\ref{sec:topo-regularization}. 
  }
  \label{TPOT-structure}
\end{figure*}
%%%%
Retinal color fundus photography (CFP) plays a critical role in diagnosing ocular diseases~\cite{deeplearning1,deeplearning5}. However, retinal fundus imaging equipment, particularly nonmydriatic cameras, faces challenges such as artifacts and blurring caused by uncontrollable factors (e.g., variability in ambient light conditions, capture errors, or operator mistakes). These issues often result in degraded retinal images. Some types of degradation, such as contrast mismatch, inconsistent illumination, and vessel occlusions, can impair the accurate visualization of lesions and blood vessels, essential for diagnosing retinal diseases such as diabetic retinopathy. Therefore, developing a robust framework to enhance low-quality images is vital for improving diagnostic accuracy\cite{zhu2023otre,zhu2023optimal}.

Since it is both difficult and costly to collect paired degraded and clean retinal images, research on retinal fundus image quality enhancement has shifted from supervised algorithms~\cite{shen2020modeling} to unsupervised algorithms, leading to remarkable improvements in image quality. These unsupervised approaches typically model retinal quality enhancement as an end-to-end image-to-image (I2I) translation task using Generative Adversarial Networks (GANs), where low-quality images $X$ are transformed into their high-quality counterparts $Y$. Specifically, CycleGAN~\cite{zhu2017unpaired} serves as a common strategy, as its cycle consistency regularization enables learning meaningful bidirectional mappings and improves generation alignment, albeit with additional computational cost and potential artifacts.

To address these limitations, Wang et al.~\cite{wang2022optimal} applied optimal transport (OT) theory to learn optimal mappings between two image domains, Zhu et al.~\cite{zhu2023otre,zhu2023optimal} introduced to use the Structural Similarity Index (SSIM) to regularize the generator and prevent excessive distortion of lesion structures. However, these methods often prove suboptimal, such as suffering from mode collapse, when handling images with complex structures or multimodal distributions. To mitigate this, Dong et al.~\cite{dong2024cunsb} proposed the unpaired neural schr\"{o}dinger bridge method to reserve the smooth and
probabilistically consistent transformation between distributions. However, their iterative learning process can smooth high-frequency information, producing suboptimal structure preservation.

Topology preservation has become a key focus in segmentation tasks. Specifically, Hu et al.~\cite{hu2019topology} extended beyond simple geometric structures by constructing a differentiable topology loss based on persistence diagrams and incorporating it into the segmentation training process. However, existing fundus image enhancement methods primarily focus on maintaining semantic consistency in the feature space~\cite{vasa2024context} without considering introducing topology preservation and leveraging the natural advantage(e.g., vessel topology structures).

In this paper, we propose a novel topology-preserving retinal fundus image enhancement framework, named \emph{Topology Preserving Optimal Transport (TPOT)}, which leverages Optimal Transport (OT) theory to learn the optimal mappings between low-quality and high-quality images while incorporating topology regularization to preserve the complex blood vessel structures. Our main contributions are threefold: (i) To the best of our knowledge, this is the first work to introduce topology-preserving into OT-GAN-based retinal enhancement methods. (ii) We extend the application of topology loss beyond segmentation tasks, demonstrating its effectiveness in preserving topological structures during image generation. (iii) We conduct experiments on public retinal fundus datasets, showing our method outperforms various supervised and unsupervised approaches in denoising and downstream vessel segmentation.
 %%%%

%%%%
\section{Topology Preserving Optimal Transport}
\label{sec:methods}
Our proposed methods contain two main modules. Optimal Transport guided domain transformation and topology-preserving regularization for retinal blood vessels. The whole framework is shown in Fig.~\ref{TPOT-structure}.

\textbf{Quality enhancement guided by optimal transport}. Let $\mathbf{X}$ and $\mathbf{Y}$ represent the domains of low-quality and high-quality images, with corresponding probability measures $\mu$ and $\nu$. The retinal fundus image enhancement task can be naturally framed as solving \textit{Monge’s} optimal transport problem, which is formulated as follows:
\begin{equation} \label{eq:ot}
    \begin{split}
        f^*  &= \inf_f \int_\mathbf{X} C(\mathbf{x}, f (\mathbf{x}) )d \mu(\mathbf{x}) \\
        &\text{subject to} \ f_{\#}\mu = \nu
    \end{split}
\end{equation}
where $C(\cdot, \cdot): \mathbf{X}\times \mathbf{Y} \rightarrow \mathbb{R}^+$ represents the cost function, $f$ denotes the candidate mapping functions, and the constraint ensures alignment between the two probability measures.
\begin{comment}
By modeling $f$ using neural networks parameterized by $\theta$, denoted as $f_\theta$, and incorporate the Lagrange multiplier $\lambda$, Eq.~\ref{eq:ot} can be relaxed into an unconstrained optimization problem as follows:
\begin{equation} \label{eq:ot-nn}
    f_{\theta^*} = \min \{ \mathbb{E}_{\mathbf{x} \sim \mathbb{P}_\mathbf{X}} [ C( \mathbf{x}, f_\theta (\mathbf{x}) )] +\lambda d(\mathbb{P}_\mathbf{Y}, \mathbb{P}_{f_\theta(\mathbf{X})} ) \}
\end{equation}
where $d(\cdot,\cdot)$ represents the difference between two image distributions $\mathbb{P}_\mathbf{X}$ and $\mathbb{P}_{f_\theta(\mathbf{x})}$, ensuring the pushforward condition.
\end{comment}
Inspired by the works in~\cite{zhu2023otre,zhu2023optimal,wang2022optimal}, Eq.~\ref{eq:ot} can be modeled using an adversarial training strategy. Specifically, the generator $G_\theta$, parameterized by $\theta$, is used to model the optimal mapping $f^*$. Identity regularization $C( \mathbf{y}, G_\theta (\mathbf{y}) )$ ensures that it doesn't introduce unnecessary changes. The discriminator $D_\beta$ helps model the Wasserstein-1 distance between $\mathbb{P}_\mathbf{Y}$ and $\mathbb{P}_{G_\theta(\mathbf{X})}$, fulfilling the pushforward condition in Eq.~\ref{eq:ot}. This leads to the following objective function: 
\begin{comment}
where the Multi-Scale Structural Similarity Index Measure (cite ssim) as the cost function $c$, and the Wasserstein-1 distance is used as the divergence function $d$. To ensure that the generator does not introduce unnecessary modifications, an additional identity regularization term, $ C( \mathbf{y}, G_\theta (\mathbf{y}) )$, is included. This leads to the following objective function:
\end{comment}

\begin{equation}
    \begin{split}
        \max_{G_\theta} \min_{D_\beta} &\mathbb{E}_{\mathbf{x} \sim \mathbb{P}_\mathbf{X}} [ C( \mathbf{x}, G_\theta (\mathbf{x}) )]  + \mathbb{E}_{\mathbf{y} \sim \mathbb{P}_\mathbf{Y}} [ C( \mathbf{y}, G_\theta (\mathbf{y}) )] + \\
        & \lambda \mathbf{W}_1(\mathbb{P}_\mathbf{Y}, \mathbb{P}_{G_\theta(\mathbf{X})})\\
    \end{split}
\end{equation}
\begin{comment}
        \text{s.t}\begin{cases}
            &\mathbf{W}_1(\mathbb{P}_\mathbf{Y}, \mathbb{P}_{G_\theta(\mathbf{X})}) = \sup_{\substack{||D_\beta||_L\leq 1}}\mathbb{E}_{\mathbf{P}_\mathbf{Y}}[D_\beta (\mathbf{Y})] -  \\
            & \hspace{3cm} \mathbb{E}_{\mathbf{P}_\mathbf{X}}[D_\beta(G_\theta(X))] \\
            &C( \mathbf{x}, G_\theta (\mathbf{x}) ) = 1- SSIM( \mathbf{x}, G_\theta (\mathbf{x}) )  \\
            & C( \mathbf{y}, G_\theta (\mathbf{y}) ) = 1-SSIM( \mathbf{y}, G_\theta (\mathbf{y}) )
        \end{cases}
\noindent Here, $G_\theta$ aims to find the optimal mapping $f_{\theta^*}$ after training, and $\mathbb{P}_\mathbf{Y}$ represents the distribution of high-quality images.
\end{comment}
\noindent Here, Cost functions $C( \mathbf{x}, G_\theta (\mathbf{x}) )$ and $C( \mathbf{y}, G_\theta (\mathbf{y}) )$ are calculated based on the locally Quasi-Convex Multi-Scale Structural Similarity Index Measure~\cite{wang2003multiscale,brunet2011mathematical}. To enforce 1-Lipschitz continuity in the discriminator, a gradient penalty is applied.
\begin{comment}
\begin{figure}[htbp]
  \centering
  \includegraphics[width=0.9\columnwidth]{images/topo_loss_illustration.pdf}
  \caption{\textbf{Left Plot:} \textbf{Col A and B}: These represent the segmentation masks of the low-quality input image and the corresponding synthetic high-quality image during training. The orange box highlights regions with complex topological structures. \textbf{Right Plot:} \textbf{Col C}: This shows the persistence diagrams, where the yellow crosses and blue dots represent corresponding persistent features in $S_N(\mathbf{x})$ and $S_N(G_\theta(\mathbf{x}))$, respectively. The green dots indicate noisy persistent features that need to be removed during training. The model penalizes the differences between corresponding points to encourage topological consistency.}
  \label{TPOT-training}
\end{figure}
\end{comment}
%%%%
\begin{table*}[!t]
\centering
\caption{Performance comparison with baselines. The best performance within each column is highlighted in bold.}
\tiny
\resizebox{0.8\textwidth}{!}{%
\begin{tabular}{cccc|cccc}
\toprule
\multirow{2}{*}{} & \multirow{2}{*}{\textbf{Method}} & \multicolumn{2}{c}{\textbf{EyeQ}}& \multicolumn{4}{c}{\textbf{Downstream vessel segmentation}} \\ \cmidrule(l){3-8} 
                                          &                                  & \textbf{SSIM} $\uparrow$   & \textbf{PSNR}$\uparrow$   & \textbf{AUC} $\uparrow$   & \textbf{PR} $\uparrow$  & \textbf{F1-score} $\uparrow$   & \textbf{SP}$\uparrow$   \\ \midrule
\multirow{1}{*}{\textit{Supervised Methods}}       & cofe-Net~\cite{shen2020modeling}                         & 0.9408           & 24.907           & 0.9188            & 0.7698            & 0.6890            & 0.9801            \\
\midrule
\multirow{8}{*}{\textit{Unsupervised Methods}}&Context-aware OT~\cite{vasa2024context} & 0.8497 & 21.088 & 0.8650 & 0.6547 &0.5887 &0.9765 \\
& CycleGAN~\cite{zhu2017unpaired}                          & 0.9313           & 25.076            & 0.9015            & 0.7277  & 0.6462            & 0.9801           \\
                                          & OTRE-GAN~\cite{zhu2023otre}                          & 0.9392           & 24.812           & 0.9156            & 0.7678           & 0.6918            & 0.9797           \\
                                          & OTT-GAN~\cite{wang2022optimal}                         & 0.9275           & 24.065           & 0.9034            & 0.7400           & 0.6608            & 0.9812  \\
                                          & WGAN~\cite{arjovsky2017wasserstein}                           & 0.9266           & 24.793           & 0.9081            & 0.7494           & 0.6768            & 0.9764           \\
                                          & CUNSB-RFIE~\cite{dong2024cunsb}                           & 0.9121           & 24.242           & 0.9163            & 0.7625           & 0.6872            & 0.9784           \\ \cmidrule(l){2-8}
                                          & Ours                       & \textbf{0.9417} & \textbf{25.196} & \textbf{0.9191}  & \textbf{0.7748}          & \textbf{0.6926}  & \textbf{0.9816}          \\ \bottomrule
\end{tabular}%
}
\label{TPOT-metrics}
\end{table*}
%%%%

\textbf{Topology-preserving regularization}.\label{sec:topo-regularization} Persistent homology provides a continuous representation of topological changes as the threshold 
$\alpha$ dynamically varies. Specifically, let $S_N$ denote the pre-trained segmentation network, $S_N(\mathbf{x})$ and $S_N(G_\theta(\mathbf{x}))$ represent the output likelihood map of low-quality input and synthetic high-quality counterparts, respectively. (e.g., $S_N(\mathbf{x})^{0.5}=\{\mathbf{x} | S_N (\mathbf{x})> 0.5 \}$ represents the binary segmentation mask for $\mathbf{x}$) As $\alpha$ gradually decreases, an increasing number of pixels in $S_N(G_\theta(\mathbf{x}))^\alpha$ are classified as foreground (e.g., blood vessels), leading to corresponding changes in the topological structures. For instance, as $\alpha$ decreases, vessels that form loops may lose their structure as more neighboring pixels are considered foreground vessels. The persistence diagrams, which summarize the changes in topology, are denoted as $ \textit{Dgm}(S_N(\mathbf{x}))$ and $\textit{Dgm}(S_N(G_\theta(\mathbf{x})))$. These diagrams yield various persistent points, denoted as $P = (d_1,b_1)$ and $P^G = (d_2,b_2)$, where $d_1$ and $d_2$ correspond to the death thresholds and $b_1$ and $b_2$ correspond to the birth thresholds of the topological features. Following the ideal outlined in~\cite{hu2019topology}, the topology regularization, $C_{topo}(\mathbf{x},G_\theta(\mathbf{x}))$, is defined as:
\begin{equation} \label{eq:topo-regu}
    \begin{split}
        C_{topo}(\mathbf{x},G_\theta(\mathbf{x})) &= \min_{\alpha}  \sum_{P^G \in \textit{Dgm}(S_N(G_\theta))} ||P^G - \alpha(P^G)||^2  \\
        =  \sum_{P^G \in \textit{Dgm}(S_N(G_\theta))}& (b_2-\alpha^*(b_2))^2 + (d_2 - \alpha^*(d_2))^2
    \end{split}
\end{equation}
where $(\alpha^*(d_2),\alpha^*(b_2)) \in P$ and $\alpha^*$ represents the optimal correspondence of points between $P$ and $P^G$. This correspondence is calculated based on the persistence (e.g., $|d_1 - b_1|$ ) of the critical points in $P$ and $P^G$. We compute the topology regularization based on image patches. The final topology regularization is expressed as:
\begin{equation}\label{eq:topo-regu-final}
    C_{topo}(\mathbf{x},G_\theta(\mathbf{x})) = \sum_i C_{topo}^i(\mathbf{x_i},G_\theta(\mathbf{x_i}))
\end{equation}
where $i$ represents the index of the image patches. As illustrated in Fig.~\ref{TPOT-structure}, by enforcing the alignment of persistent dots in $P^G$ with their corresponding dots in $P$ and by removing noisy dots that lack matches, we ensure that the enhancement process preserves the structural integrity of blood vessels. Consequently, as training progresses, the complex topological structure of the blood vessels (highlighted in the orange box) gradually becomes more consistent between $\mathbf{x}$ and $G_\theta(\mathbf{x})$. Additionally, since Eq.~\ref{eq:topo-regu} is computed based on the mean squared error between critical points in $P$ and $P^G$, this regularization is convex and has a well-established gradient~\cite{hu2019topology}. Thus, the final training objective is given by:
\begin{equation}\label{eq:final-object}
    \begin{split}
        &\max_{G_\theta} \min_{D_\beta} \lambda_{ssim} \mathbb{E}_{\mathbf{x} \sim \mathbb{P}_\mathbf{X}} [ C( \mathbf{x}, G_\theta (\mathbf{x}) )] +  \\
        &\lambda_{idt}\mathbb{E}_{\mathbf{y} \sim \mathbb{P}_\mathbf{Y}} [ C( \mathbf{y}, G_\theta (\mathbf{y}) )] +  \\ 
        &\lambda_{topo} \mathbb{E}_{\mathbf{x} \sim \mathbb{P}_\mathbf{X}} [  C_{topo}( \mathbf{x}, G_\theta (\mathbf{x}) )] \mathbb{I}_{t_i} + \\
        &\mathbf{W}_1(\mathbb{P}_\mathbf{Y}, \mathbb{P}_{G_\theta(\mathbf{X})})
    \end{split}
\end{equation}
\noindent where $\mathbb{I}_{t_i}$ is the indicator function, signifying that $G_\theta$ incorporates topology regularization after $t_i$ epochs.

\begin{figure*}[!t]
  \centering
  \includegraphics[width=1.0\textwidth]{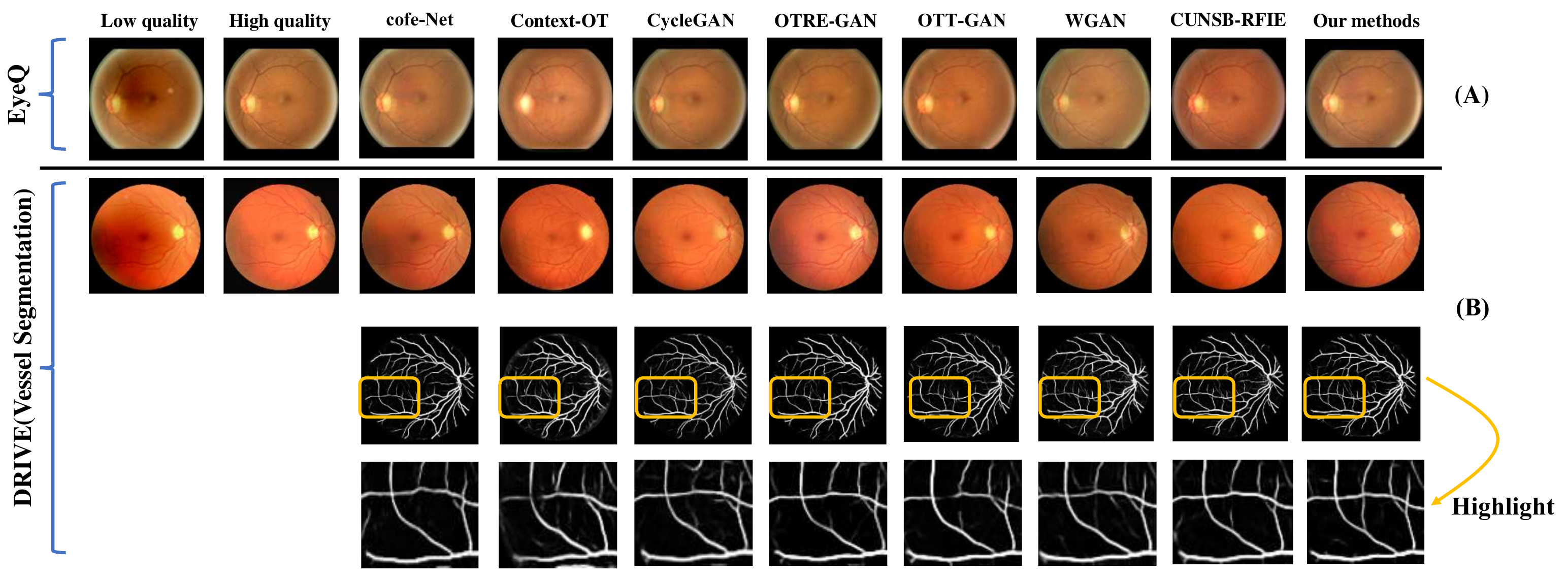}  % Replace with your figure file and remove the example
  \vspace{-0.4cm}
  \caption{\textbf{(A)} illustrates the result of quality enhancement task on the EyeQ dataset. \textbf{(B)} shows the enhanced DRIVE image along with the corresponding vessel segmentation. The orange box highlights a patch with a complex topological structure, where our method demonstrates greater consistency.
  }
  \label{TPOT-result-ill}
\end{figure*}
%%%
\begin{comment}
\begin{table}[ht]
\centering
\tiny  % Use \scriptsize for smaller font
\resizebox{0.75\columnwidth}{!}
{
%  % Adjust width to better fit the column
\begin{tabular}{ccc}
\hline
$\lambda_{topo}$  & \textbf{SSIM} & \textbf{PSNR} \\ \hline
$0$ (OTE-GAN)  & 0.9392 & 24.812 \\
$0.1$ &  0.9409 & 24.963 \\
$1$ &  0.9417 & 25.196 \\
$10$ & 0.9413 & 25.126 \\ \hline
\end{tabular}
}
\caption{Ablation study over Topology-preserving regularization.}
\label{tab:ablation}
\end{table}
\end{comment}
%%%
\section{Experiments and results}
\label{sec:experiments}
\subsection{Experimental Details}
\noindent \textbf{Baselines}
We compare our methods against the following baselines: \textit{Supervised algorithm}: cofe-Net \cite{shen2020modeling}; \textit{Unsupervised algorithms}:  CycleGAN~\cite{zhu2017unpaired}, OTT-GAN~\cite{wang2022optimal}, OTRE-GAN~\cite{zhu2023otre}, WGAN~\cite{arjovsky2017wasserstein}, Context-aware OT~\cite{vasa2024context} and CUNSB-RFIE~\cite{dong2024cunsb}. For the quality enhancement task, we utilize Structural Similarity Index Measure (SSIM) and Peak Signal-to-Noise Ratio (PSNR) as evaluation metrics. For the downstream vessel segmentation task, we assess performance using Area under the ROC Curve (AUC), Area under the Precision-Recall Curve (PR), F1-score and Specificity (SP).

\noindent \textbf{Dataset}. For the quality enhancement task, we used the algorithms outlined in~\cite{shen2020modeling} to generate degraded images that combined spot artifacts, illumination changes, and blurring. All models were trained on 5,000 EyeQ images~\cite{eyeQ} and tested on 6,217 EyeQ images. The training utilized both supervised and unsupervised modes, depending on the specific methods employed. The vessel segmentation task was conducted using the enhanced images, starting from scratch with the DRIVE~\cite{1282003} dataset, adhering to the official training and testing split. Before any operations, All images are cropped and resized into $256 \times 256$ pixels, except method~\cite{shen2020modeling}.

\noindent\textbf{TPOT settings}. The generator and discriminator used in Eq.~\ref{eq:final-object} follows the design outlined in~\cite{zhu2023otre,zhu2023optimal}, and the pre-trained segmentation network employed from~\cite{zhou2021study}. We keep the same parameter settings with~\cite{shen2020modeling, zhu2023optimal,vasa2024context,dong2024cunsb} for all baselines. For TPOT, we trained the model for 200 epochs using the RMSprop optimizer, with $t_i = 100$. The learning rate in each phase started at $2 \times 10^{-4}$ and decayed by a factor of 10 after every 50 epochs. Additional parameters were set as follows: $\lambda_{ssim}=30,\lambda_{idt}=15,\lambda_{topo}=1$ and the patch size in Eq.~\ref{eq:topo-regu-final} was set to 65.

\subsection{Experimental Results}
\textbf{Quality enhancement task}. The metric scores are shown in Tab.~\ref{TPOT-metrics}, where our methods achieve the best performance in both SSIM and PSNR, highlighting its superiority in preserving perceptual quality and pixel-wise alignment. For other models, as shown in Fig.~\ref{TPOT-result-ill}\textbf{(A)}, the Context-aware OT method shows noticeable semantic deficiencies at the image boundaries, resulting in the disappearance of blood vessels. In contrast, CUNSB-RFIE maintains better consistency in image space and effectively handles spot artifacts, though it slightly alters the background. Other GAN-based models (e.g., CycleGAN, OTRE-GAN, OTT-GAN) struggle to preserve peripheral blood vessels with complex topological structures (e.g., loops), still resulting in over-tampering of vessel structures. Our method outperforms these approaches by preserving overall image quality and vessel structures. 

\noindent \textbf{Vessel segmentation task}. As shown in Tab.~\ref{TPOT-metrics}, our method achieves the best performance across all four evaluation metrics. Specifically, Fig.~\ref{TPOT-result-ill}\textbf{(B)} presents the vessel segmentation illustration on the DRIVE dataset. Due to vessel deficiencies at the image level, the Context-OT model produces noisy boundary predictions. Other GAN-based methods (e.g., OTRE-GAN and CycleGAN) improve overall image quality but struggle to preserve peripheral vessel structures. In the highlighted segmentation patch, CUNSB demonstrates a better ability to predict small vessel branches, while other methods tend to classify them as background. Our method preserves the complex topology structure, leading to continuous and consistent vessel morphology. 

\begin{table}[ht]
\centering
\tiny  % Use \scriptsize for smaller font
\resizebox{0.75\columnwidth}{!}
{
%  % Adjust width to better fit the column
\begin{tabular}{ccc}
\hline
$\lambda_{topo}$  & \textbf{SSIM} & \textbf{PSNR} \\ \hline
$0$ (OTRE-GAN)  & 0.9392 & 24.812 \\
$0.1$ &  0.9409 & 24.963 \\
$1$ &  0.9417 & 25.196 \\
$10$ & 0.9413 & 25.126 \\ \hline
\end{tabular}
}
\caption{Ablation study over Topology-preserving regularization.}
\label{tab:ablation}
\end{table}
\subsection{Ablation Study}
Our method leverages topology-preserving regularization to maintain the blood vessel structure during enhancement. To evaluate the effect of this additional regularization on the final generation performance, we conducted ablation studies on the EyeQ dataset, with the results reported in Tab.~\ref{tab:ablation}. The absence of topology-preserving regularization, as represented by OTRE-GAN, resulted in relatively poorer performance. As we incrementally increased the weight $\lambda_{topo}$ in Eq.~\ref{eq:final-object}, we observed a clear performance improvement, even with a small weight. This demonstrates that topology-preserving regularization contributes significantly to the overall enhancement quality.

\section{Conclusion}
\label{sec:conclusion}

In this study, we propose a Topology Preserving Optimal Transport (TPOT) pipeline for enhancing retinal images. We incorporate topology regularization to preserve blood vessel structures, promoting vessel alignment in the enhancement process. We demonstrate the effectiveness of TPOT in denoising and maintaining the topological integrity of blood vessels on EyeQ and DRIVE datasets, respectively. While we utilized synthetic degraded images and faced challenges related to domain shifts in capturing segmentation outputs, our findings underscore the importance of preserving topological structures in medical image generation. Future research will further explore the potential of topology regularization in other medical image generation applications.

\section{ COMPLIANCE WITH ETHICAL STANDARDS}
This research study was conducted retrospectively using human subject data available in open access by~\cite{eyeQ}. Ethical approval was not required as confirmed by the license attached with the open-access data.
\section{ACKNOWLEDGMENTS}
This work was supported by grants from the National Institutes of Health (RF1AG073424, R01EY032125, and R01DE0\\30286) and the State of Arizona via the Arizona Alzheimer Consortium.


% ------------------------------------------------------------------------- 
\bibliographystyle{IEEEbib}
\bibliography{strings,refs}

\begin{thebibliography}{10}

\bibitem{zhu2023otre}
Wenhui Zhu, Peijie Qiu, Oana~M Dumitrascu, Jacob~M Sobczak, Mohammad Farazi, Zhangsihao Yang, Keshav Nandakumar, and Yalin Wang,
\newblock ``Otre: Where optimal transport guided unpaired image-to-image translation meets regularization by enhancing,''
\newblock in {\em International Conference on Information Processing in Medical Imaging}. Springer, 2023, pp. 415--427.

\bibitem{zhu2023optimal}
Wenhui Zhu, Peijie Qiu, Mohammad Farazi, Keshav Nandakumar, Oana~M Dumitrascu, and Yalin Wang,
\newblock ``Optimal transport guided unsupervised learning for enhancing low-quality retinal images,''
\newblock {\em Proc IEEE Int Symp Biomed Imaging}, 2023.

\bibitem{zhou2021study}
Yuqian Zhou, Hanchao Yu, and Humphrey Shi,
\newblock ``Study group learning: Improving retinal vessel segmentation trained with noisy labels,''
\newblock in {\em Medical Image Computing and Computer Assisted Intervention--MICCAI 2021: 24th International Conference, Strasbourg, France, September 27--October 1, 2021, Proceedings, Part I 24}. Springer, 2021, pp. 57--67.

\bibitem{deeplearning1}
Wenhui Zhu, Peijie Qiu, Xiwen Chen, Xin Li, Natasha Lepore, Oana~M. Dumitrascu, and Yalin Wang,
\newblock ``nnmobilenet: Rethinking cnn for retinopathy research,''
\newblock in {\em Proceedings of the IEEE/CVF Conference on Computer Vision and Pattern Recognition (CVPR) Workshops}, June 2024, pp. 2285--2294.

\bibitem{deeplearning5}
Oana~M. Dumitrascu, Xin Li, Wenhui Zhu, Bryan~K. Woodruff, Simona Nikolova, Jacob Sobczak, Amal Youssef, Siddhant Saxena, Janine Andreev, Richard~J. Caselli, John~J. Chen, and Yalin Wang,
\newblock ``Color fundus photography and deep learning applications in alzheimer’s disease,''
\newblock {\em Mayo Clinic Proceedings: Digital Health}, 2024.

\bibitem{shen2020modeling}
Ziyi Shen, Huazhu Fu, Jianbing Shen, and Ling Shao,
\newblock ``Modeling and enhancing low-quality retinal fundus images,''
\newblock {\em IEEE transactions on medical imaging}, vol. 40, no. 3, pp. 996--1006, 2020.

\bibitem{zhu2017unpaired}
Jun-Yan Zhu, Taesung Park, Phillip Isola, and Alexei~A Efros,
\newblock ``Unpaired image-to-image translation using cycle-consistent adversarial networks,''
\newblock in {\em Proceedings of the IEEE international conference on computer vision}, 2017, pp. 2223--2232.

\bibitem{wang2022optimal}
Wei Wang, Fei Wen, Zeyu Yan, and Peilin Liu,
\newblock ``Optimal transport for unsupervised denoising learning,''
\newblock {\em IEEE Transactions on Pattern Analysis and Machine Intelligence}, vol. 45, no. 2, pp. 2104--2118, 2022.

\bibitem{dong2024cunsb}
Xuanzhao Dong, Vamsi~Krishna Vasa, Wenhui Zhu, Peijie Qiu, Xiwen Chen, Yi~Su, Yujian Xiong, Zhangsihao Yang, Yanxi Chen, and Yalin Wang,
\newblock ``Cunsb-rfie: Context-aware unpaired neural schr"$\{$o$\}$ dinger bridge in retinal fundus image enhancement,''
\newblock {\em arXiv preprint arXiv:2409.10966}, 2024.

\bibitem{hu2019topology}
Xiaoling Hu, Fuxin Li, Dimitris Samaras, and Chao Chen,
\newblock ``Topology-preserving deep image segmentation,''
\newblock {\em Advances in neural information processing systems}, vol. 32, 2019.

\bibitem{vasa2024context}
Vamsi~Krishna Vasa, Peijie Qiu, Wenhui Zhu, Yujian Xiong, Oana Dumitrascu, and Yalin Wang,
\newblock ``Context-aware optimal transport learning for retinal fundus image enhancement,''
\newblock {\em arXiv preprint arXiv:2409.07862}, 2024.

\bibitem{wang2003multiscale}
Zhou Wang, Eero~P Simoncelli, and Alan~C Bovik,
\newblock ``Multiscale structural similarity for image quality assessment,''
\newblock in {\em The Thrity-Seventh Asilomar Conference on Signals, Systems \& Computers, 2003}. Ieee, 2003, vol.~2, pp. 1398--1402.

\bibitem{brunet2011mathematical}
Dominique Brunet, Edward~R Vrscay, and Zhou Wang,
\newblock ``On the mathematical properties of the structural similarity index,''
\newblock {\em IEEE Transactions on Image Processing}, vol. 21, no. 4, pp. 1488--1499, 2011.

\bibitem{arjovsky2017wasserstein}
Martin Arjovsky, Soumith Chintala, and L{\'e}on Bottou,
\newblock ``Wasserstein generative adversarial networks,''
\newblock in {\em International conference on machine learning}. PMLR, 2017, pp. 214--223.

\bibitem{eyeQ}
Huazhu Fu, Boyang Wang, Jianbing Shen, and et~al.,
\newblock ``Evaluation of retinal image quality assessment networks in different color-spaces,''
\newblock {\em MICCAI}, pp. 48--56, 2019.

\bibitem{1282003}
J.~Staal, M.D. Abramoff, M.~Niemeijer, M.A. Viergever, and B.~van Ginneken,
\newblock ``Ridge-based vessel segmentation in color images of the retina,''
\newblock {\em IEEE Transactions on Medical Imaging}, vol. 23, no. 4, pp. 501--509, 2004.

\end{thebibliography}

\end{document}